\documentclass[10pt]{iopart}
\usepackage{iopams}
\usepackage{epsfig}
\DeclareFontFamily{OT1}{rsfs}{}
\DeclareFontShape{OT1}{rsfs}{m}{n}{<5> rsfs5 <7> rsfs7 <10>
rsfs10}{}
\DeclareSymbolFont{mathrsfs}{OT1}{rsfs}{m}{n}
\DeclareSymbolFontAlphabet{\mathrsfs}{mathrsfs}
\begin{document}
\paper{Unified \textit{ab initio} treatment of attosecond photoionization and Compton scattering}
\author{G~L~Yudin$^{1, 2}$, D~I~Bondar$^{2,3}$,    S~Patchkovskii$^{2}$, P~B~Corkum$^{2,4}$ and A~D~Bandrauk$^{1}$}
\address{$^{1}$ Universit\'e de Sherbrooke, Sherbrooke, Qu\'ebec J1K 2R1, Canada}
\address{$^{2}$ National Research Council of Canada, Ottawa, Ontario K1A 0R6, Canada}
\address{$^{3}$ University of Waterloo, Waterloo, Ontario N2L 3G1, Canada}
\address{$^{4}$ University of Ottawa, Ottawa, Ontario K1N 6N5, Canada}

\ead{gennady.yudin@nrc.ca}
\begin{abstract}
We present a new theoretical approach to attosecond laser-assisted photo- and Compton ionization. \textit{Attosecond} x-ray absorption and scattering are described by  $\hat{\mathrsfs{S}}^{(1,2)}$-matrices, which are coherent superpositions of \textit{``monochromatic''} $\hat{S}^{(1,2)}$-matrices in a laser-modified Furry representation. Besides refining the existing theory of the soft x-ray photoelectron attosecond streak camera and spectral phase interferometry (ASC and ASPI), we formulate a theory of hard x-ray photoelectron and Compton ASC and ASPI. The resulting scheme has a simple structure and leads to closed-form expressions for ionization amplitudes. We investigate Compton electron interference in the separable Coulomb-Volkov continuum with both Coulomb and laser fields treated non-perturbatively. We find that at laser-field intensities below 10$^{13}$ Wcm$^{-2}$ normalized Compton lines almost coincide with the lines obtained in the laser-free regime. At higher intensities, attosecond interferences survive integration over electron momenta, and feature prominently in the Compton lines themselves. We define a regime where the electron ground-state density can be measured with controllable accuracy in an attosecond time interval. The new theory provides a firm basis for extracting photo- and Compton electron phases and atomic and molecular wavefunctions from experimental data.
\end{abstract}
%





%
\section{Introduction}
Attosecond science began in the 21st century starting with the experiment \cite{Paul2001} where a train of 250 attosecond radiation pulses was obtained, and the measurement \cite{Hentschel2001} which indicated a soft-x-ray pulse duration of 650$\pm$150 attoseconds (asec). Fundamental to the generation process, the attosecond pulses are synchronized with the infrared field. This precise synchronization between the \textit{soft x-ray} attosecond pulse and an intense laser field is prominent in attosecond science. The infrared field can provide the nonlinearity needed for measurement of attosecond pulses \cite{Itatani2002,Kitzler2002,Quere2003,Itatani2004} and investigation of attosecond phenomena in physics, chemistry and technology \cite{Corkum2007}. The current state of the theory and experimental techniques in attosecond metrology is summarized in recent reviews \cite{Krausz2009,Nisoli2009}.

Recent developments in ultrafast \textit{hard x-ray} sources demonstrate that it is technically feasible to generate \textit{hard x-ray attosecond pulses in the {\AA}ngstr\"{o}m wavelength range} \cite{Saldin2002,Zholents2004,Zholents2005a,Zholents2005b,Zholents2008,Ding2009} with free-electron lasers (FELs). At the very heart of these proposals is the temporal synchronization scheme between the FEL and an external source---the so-called seeded attosecond x-ray radiation \cite{Zholents2004}.

The radiation of a relativistic electron interacting with a co-propagating tightly focused high-power laser has also been investigated \cite{Kim2009}. This scheme can be used for attosecond/femtosecond source in the hard x-ray region of 10--100 keV.

Besides the laser-assisted attosecond photoionization, other hard x-ray radiation is involved in physical phenomena exhibiting dynamic response on the attosecond time scale, which has not been explored so far. We analyze one such process---Compton ionization. The relative contributions of the photo- and Compton ionization processes are sensitive to the incident photon energy $\hbar \Omega$ and ionization potential $I_{p}$. In the case of a weakly bound electron ($I_{p}\ll \hbar \Omega$), Compton ionization prevails at incident photon energies of \cite{Bethe1957}
\begin{eqnarray} \label{Photo-Compton}
\hbar \Omega \gtrsim 3(Z_{a})^{8/7} keV,
\end{eqnarray}
where $Z_{a}$ denotes the exponential coefficient in the initial electronic wavefunction near nucleus\footnote{Unless stated otherwise, the Hartree atomic units (a.u., $|e|=m_{e}=\hbar=1$ and the velocity of light $c\approx$137.036) are used below.} $| i\rangle \varpropto \exp{(-Z_{a}r)}$.

Soon after Compton's discovery \cite{Compton1923}, Du Mond \cite{Du Mond1929} suggested an explanation of Compton scattering by bound electrons. He proposed the so-called impulse approximation: a bound electron is assumed to be ``nearly'' free, and its initial momentum distribution gives rise to broadening of Compton lines. Eisenberger and Platzmann \cite{Eisenberger1970} developed the non-relativistic quantum-mechanical theory of Compton profiles, thus justifying the Du Mond's hypothesis.

Recent advent of high-brilliance synchrotron radiation sources turned Compton scattering experiments into a \textit{powerful direct probe} of the electronic and magnetic structure of materials \cite{Cooper2004}. Its high precision permits accurate measurements of the electron momentum density of atoms, molecules and extended solids, stimulating developments of the theory of many-body systems. The three-dimensional electron momentum density can be either reconstructed from the measured Compton profiles, or acquired directly with the ($\gamma$,e$\gamma$) technique---the coincident detection of the scattered x-ray photon and the recoil electron \cite{Schulke2007}.

With the practical attosecond hard x-ray sources becoming available, it is essential to generalize the Compton profile theory to the attosecond regime. It is also necessary to extend the ideas of ASC and ASPI to the hard x-ray range. Transferring these techniques to laser-assisted Compton ionization is far from trivial. The photon is scattered, not absorbed, by the weakly bound electron, leading to a quite distinct theoretical description. In spite of these complications we will show that the Compton electron contains complete phase information on the incident pulse provided that we select the momentum of the scattered photon.

We formulate an analytic theory of laser-assisted attosecond hard x-ray absorbtion and Compton scattering. We concentrate on Compton ionization, as a relatively more complicated process. To obtain the appropriate semi-relativistic scattering matrix, we generalize the theory \cite{Akhiezer1969} of monochromatic photon scattering by a bound electron and incorporate the separable Coulomb-Volkov continuum \cite{Yudin2007,Yudin2008} into the Furry representation \cite{Furry1951}. Using hydrogen atom as an example, we analyze the Compton ASC and ASPI and demonstrate the feasibility of measuring Compton profiles in attosecond laser-assisted regime. We show that Compton electron spectra depend on the phase of the laser field relative to the x-ray attosecond pulse, just as in the familiar photoelectron case \cite{Itatani2002,Kitzler2002,Quere2003,Itatani2004}. Thus, the relative phase of all frequency components of the attosecond pulse can be determined from the spectrum of Compton electrons. Finally, we analyze laser-assisted Compton lines and define a regime where the linear relationship holds between the \textit{attosecond} Compton lines and \textit{monochromatic} Compton profiles.
\section{Theory}
\subsection{Laser-free Compton scattering}
As is well established in traditional quantum electrodynamics, a general solution of Maxwell equations can be represented as a superposition of plane waves. If electromagnetic field is confined to a cubic box of volume $V$, the photon momentum $\mathbf{Q}$ can have only discrete values, $Q_{n}\sim n V^{-1/3}$ ($n=0,\pm$ 1,$\pm$ 2,...). The set $Q_{n}$ approaches a continuum when $V \rightarrow \infty$; thus, the sum over $k_{n}$ transforms into a three-dimensional Fourier integral. To describe the potential of the initial (\textit{attosecond}) photon state, we employ the following general representation:
\begin{eqnarray} \label{atto-pulse}
\mathbf{A}_{1}(\mathbf{r},t)=\mathbf{e}_{1}
\int d^{3}q \int_{-\infty}^{\infty} d\omega \frac{ F(\mathbf{q},\omega)}{\sqrt{\Omega_{1}'}}
e^{i\mathbf{k}_{1}'\mathbf{r} -i \Omega_{1}' t },
\end{eqnarray}
where $\mathbf{e}_{1}$ is the polarization vector, $\mathbf{k}_{1}'$ and $\Omega_{1}'$ are momenta and energies of the incoming photons respectively, $\Omega_{1}'=\Omega_{1}+\omega$, $\Omega_{1}$ is the central frequency, $\mathbf{k}_{1}'=\mathbf{k}_{1}+\mathbf{q}$ and
$k_{1}'=\Omega_{1}'/c$. Without loss of generality, we fix the momentum of the scattered photon (the direction and frequency), so that the potential of the final photon state is represented by a plane wave:
\begin{eqnarray} \label{scatt-photon}
\mathbf{A}_{2}(\mathbf{r},t)=\mathbf{e}_{2}c
\sqrt{\frac{2\pi}{V \Omega_{2}}}
e^{i\mathbf{k}_{2}\mathbf{r} -i \Omega_{2}t},
\end{eqnarray}
where $\mathbf{k}_{2}$ and $\Omega_{2}$ are the momentum and
energy of the outgoing photon and $k_{2}=\Omega_{2}/c$.

Now we define the generally unknown function $F(\mathbf{q},\omega)$. We confine ourselves to the experimental setup where the direction of the momentum $\mathbf{k}_{1}'$ is fixed. In this geometry, equation (\ref{atto-pulse}) contains only a one-dimensional integral over $q$. We utilize the well-established Furry representation \cite{Furry1951}, used previously for the analysis of monochromatic photon scattering by a bound electron \cite{Akhiezer1969}, to obtain
the $\hat{\mathrsfs{S}}^{(2)}$-matrix element describing \textit{attosecond} Compton scattering:
\begin{eqnarray} \label{S2-matrix}
\mathrsfs{S}^{(2)}_{fi} &\propto& \int d q \int d t_{1} \int
d t_{2} \int
d\omega  \int d \nu \frac{F(q,\omega)}{\sqrt{\Omega_{1}'}} \nonumber \\
&\times&
[e^{i(E_{f}+\Omega_{2}+\nu)t_{2}-i(E_{i}+\Omega_{1}'+\nu)t_{1}} S_{1}(\nu)
\nonumber \\
&+& e^{i(E_{f}-\Omega_{1}'+\nu)t_{2}-i(E_{i}-\Omega_{2}+\nu)t_{1}}
S_{2}(\nu)],
\end{eqnarray}
\begin{eqnarray} \label{Summa1}
S_{1}(\nu)= \sum_{s}
\frac{\langle
f\mid \hat{b}_{2}\mid s\rangle
\langle s\mid \hat{b}_{1}\mid
i\rangle}{E_{s}+\nu},
\end{eqnarray}
\begin{eqnarray} \label{Summa2}
S_{2}(\nu)= \sum_{s}
\frac{\langle
f\mid \hat{b}_{1}\mid s\rangle
\langle s\mid \hat{b}_{2}\mid
i\rangle}{E_{s}+\nu}.
\end{eqnarray}
where $\hat{b}_{1}= \hat{e}_{1}e^{i\mathbf{k}_{1}'\mathbf{r}}$, $\hat{b}_{2}=\hat{e}_{2}e^{-i\mathbf{k}_{2}\mathbf{r}}$, $\hat{e}_{1,2}=\hat{\overrightarrow{\gamma}}\cdot
\mathbf{e}_{1,2}$, $\hat{\gamma}_{\mu}$ are the Dirac matrices ($\mu$ = 1, 2, 3), $|s\rangle$ are the intermediate electron states with energies $E_s$ and  $| i\rangle$ and $| f\rangle$ are the initial and final states with the energies $E_{i}$ and $E_{f}$ respectively. The \textit{``attosecond''} Compton $\hat{\mathrsfs{S}}^{(2)}$-matrix describing the laser-free regime is a coherent superposition [through the integrals over q and $\omega$ in formula (\ref{S2-matrix})] of the \textit{``monochromatic''} $\hat{S}^{(2)}$-matrices in the Furry picture. The matrix $\hat{S}^{(2)}$ is represented by upper diagrams in figure 1. Labels 1 and 2 in figure 1 correspond to the initial and final electron and photon states respectively. The four-dimensional vectors $x_{1,2}$ are $(\mathbf{r}_{1,2},ict_{1,2})$. Dotted lines correspond to photon states, and solid lines describe electron states. $\psi_{1,2}(x_{1,2})$ and $S_{c}^{(e)}(x_{1},x_{2})$ are the electron wavefunctions and Green's function.

Integration in the $\mathrsfs{S}^{(2)}_{fi}$-matrix element leads to $\delta$-functions, one of which fixes the parameter $\omega$. Having fixed $\omega$, we also set the momentum $\mathbf{k}_{1}'$, i.e. remove the integral over $q$. Now we select the initial x-ray pulse as a sum of linearly chirped Gaussians with a characteristic temporal duration $\tau$ (cf with the case of attosecond photoionization \cite{Yudin2007,Yudin2008}):
\begin{eqnarray} \label{atto-pulse-2}
\mathbf{A}_{1}(\mathbf{r},t)= \mathbf{e}_{1} D c \sqrt{V}
\int_{-\infty}^{\infty} \frac{R(\omega)}{\sqrt{\Omega_{1}'}}
 e^{i\mathbf{k}_{1}'\mathbf{r} -i \Omega_{1}' t }d\omega,
\end{eqnarray}
where the general attosecond interference factor $R(\omega)$ is determined by the replica weights $\zeta_{k}$ as
\begin{eqnarray} \label{R-factor}
R(\omega)=e^{-(1-i\xi) \omega^{2} \tau^{2} /2}  \sum_{k}
\zeta_{k}e^{i\omega t_{k}},
\end{eqnarray}
where $t_{k}$ is time of the replica maxima, $D$ is $V^{-1}A_{0}\tau
\sqrt{1-i\xi}$, $A_{0}$ is the amplitude of the pulse and $\xi$ is the dimensionless chirp. Equations (\ref{atto-pulse-2}) and (\ref{R-factor}) describe a set of attosecond pulses with the bandwidth invariant to the dimensionless chirp $\xi$. The individual pulse duration is $\tau_{FWHM}=2\tau \sqrt{\ln{2}}
\sqrt{1+\xi^{2}}$.

Starting with equations (\ref{scatt-photon})--(\ref{atto-pulse-2}), we obtain the following expression for the $\hat{\mathrsfs{S}}^{(2)}$-matrix element which describes attosecond Compton scattering:
\begin{eqnarray} \label{S-matrix-1}
\mathrsfs{S}^{(2)}_{fi} &=& -(2\pi)^{5/2} i D R(\omega)U_{fi}^{(2)}, \end{eqnarray}
\begin{eqnarray} \label{U-matrix}
U_{fi}^{(2)}= (\Omega_{1}'\Omega_{2})^{-1/2}[S_{1}(\nu_{1})+S_{2}(\nu_{2})],
\end{eqnarray}
with $\nu_{1}=-E_{i}-\Omega_{1}'$, $\nu_{2}=-E_{i}+\Omega_{2}$ and $\omega =E_{f} + \Omega_{2}- E_{i}-\Omega_{1}$. We will further consider semi-relativistic Compton ionization
wherein
\begin{eqnarray} \label{omega-1}
\omega \equiv \omega_{1}=p^{2}/2+I_{p}+\Omega_{2}-\Omega_{1}
\end{eqnarray}
and $\mathbf{p}$ is the asymptotic value of Compton electron momentum at infinity. A closed-form expression is available for the transition matrix element, provided that weakly bound electron, $I_{p}/\Omega_{1}'\ll 1$, semi-relativistic photon energies, $\Omega_{1}'/c^{2}\ll 1$, and $Z_{a}/c \ll 1$ are assumed. The matrix elements in
the numerators of $U_{fi}$ determine the degree of contribution of different intermediate states $| s\rangle$ to sums (\ref{Summa1}) and (\ref{Summa2}). If the initial electronic state with the effective charge $Z_{a}$ is non-relativistic, $p_{i}\sim Z_{a}\ll c$, then only the intermediate states with non-relativistic momenta and energies, $p_{s}\lesssim Z_{a}+\Omega_{1}'/c$ and $|\, | E_{s} | - c^{2}| \lesssim I_{p}+(\Omega_{1}'/c)^{2}$, contribute to the amplitude $U_{fi}$. Note that one should neglect the second-order terms with respect to the small parameters $Z_{a}/c$,
$\Omega_{1}'/c^{2}$ and $I_{p}/\Omega_{1}'$,
e.g. $(\Omega_{1}'-\Omega_{2})/c^{2}$. However,  such a
first-order approximation is not valid on the wings of the Compton line. Furthermore, one can also neglect
the terms $\mid E_{i,f}- c^{2}\mid/c^{2}\sim
(Z_{a}/c)^{2}$ and $\mid \mid E_{s}\mid - c^{2}\mid /c^{2} \sim (\Omega_{1}'/c^{2})^{2}$. Following the approach \cite{Yudin1976} originally developed for laser-free monochromatic Compton scattering (further details can also be found in \cite{Dykhne1996}), one can obtain a first-order approximation for the $U_{fi}^{(2)}$-matrix:
\begin{eqnarray} \label{ampl-comp}
U_{fi}^{(2)}\approx \mathfrak{M}_{fi}^{(2)}(\mathbf{p})= \langle f |
\exp{(i\mathbf{k}\mathbf{r})[\hat{G}_{0}+\hat{G}_{1}+\hat{G}_{2}]}
| i\rangle ,
\end{eqnarray}
where the following notation is employed:
\numparts \begin{eqnarray} \label{G-0-2}
\hat{G}_{0}= (\Omega_{1}'\Omega_{2})^{-1/2}(\mathbf{e}_{1}\cdot
\mathbf{e}_{2}), \\
\hat{G}_{1}=(\Omega_{1}')^{-2}[(\mathbf{e}_{2}\cdot
\mathbf{k}_{1}')(\mathbf{e}_{1}\cdot\hat{\mathbf{p}})
+(\mathbf{e}_{1}\cdot \mathbf{k}_{2})(\mathbf{e}_{2}\cdot
\hat{\mathbf{p}})], \\
\hat{G}_{2}=(1/2c^{2}) \{ ([\mathbf{e}_{1}\times\mathbf{e}_{2}]\cdot
\hat{\overrightarrow{\gamma}})
\hat{\gamma_{1}}\hat{\gamma_{2}}\hat{\gamma_{3}} \},
\end{eqnarray}
\endnumparts
$\hat{\mathbf{p}}=-i\overrightarrow{\nabla}$ is the non-relativistic momentum operator and the vector
$\mathbf{k}=\mathbf{k}_{1}'-\mathbf{k}_{2}$. Terms $\hat{G}_{0}$ and $\hat{G}_{2}$ in equation (\ref{ampl-comp}) correspond to the sum over intermediate states $| s\rangle$ with negative energies (the positronic states: $E_{s}<0$), while $\hat{G}_{1}$ corresponds to the sum over intermediate states $| s\rangle$ with positive energies (the electronic states: $E_{s}>0$). It should be noted that the amplitude $\mathfrak{M}_{fi}^{(2)}(\mathbf{p})$ can be used to study bound-bound, bound-free and free-free electronic transitions.
\subsection{Laser-assisted photo- and Compton ionization}
Treating the IR laser field classically in the dipole approximation, we define its vector potential as
\begin{eqnarray} \label{vect-pot-las}
\mathbf{A}_{L}(t)= -\frac{c}{\omega_{0}}\Bigl[\mathbf{E}_{1}\sin
(\omega_{0}t)+\mathbf{E}_{2}\cos (\omega_{0}t)\Bigr]e^{-\delta
|t|},
\end{eqnarray}
where $\omega_{0}$ is the laser-field frequency, the electrical field component $\mathbf{E}_{1}$ is along the Cartesian $z$ axis, while the vector $\mathbf{E}_{2}$ is along the $x$ axis. The vector potential $\mathbf{A}_{L}(t)$ and electric field $\mathbf{E}_{L}(t)=-(1/c)d\mathbf{A}_{L}(t)/ dt$ are
turned on adiabatically at $t\rightarrow -\infty$ and turned off at $t\rightarrow\infty$ (i.e., $\delta \rightarrow +0$).

For Compton ionization, we consider the semi-relativistic regime (\ref{ampl-comp}) with electron transition from a laser-free bound state to the separable Coulomb-Volkov continuum \cite{Yudin2007}. Accordingly, the "attosecond" Compton $\hat{\mathrsfs{S}}^{(2)}$ matrix is a coherent superposition of $\hat{S}^{(2)}$ matrices describing laser-assisted monochromatic x-ray Compton ionization. The corresponding $\hat{S}^{(2)}$ matrix is represented by the lower, laser-modified ``seagull'' diagram in figure 1. For photoionization, we consider the non-relativistic regime.

The three-fold differential photoelectron spectrum and the six-fold differential probability of Compton ionization can be written as
\begin{eqnarray} \label{spectrum-photo}
\frac{d^{3}P_{Photo}}{dpdO_{e}} =
\Bigl(\frac{4\pi^{2}p}{c}\Bigr)^{2}V\mid
\mathfrak{M}_{atto}^{(1)}(\mathbf{p}) \mid^{2},
\end{eqnarray}
\begin{eqnarray} \label{spectrum-6}
\frac{d^{6}P_{Comp}}{dpdO_{e}d\Omega_{2}dO_{2}}\equiv
S^{(6)}_{Comp} = \frac{ (2\pi p \Omega_{2})^{2}}{c^{3}}
|\mathfrak{M}_{atto}^{(2)}(\mathbf{p}) |^{2},
\end{eqnarray}
where $dO_{e}=\sin\theta_{e} d\theta_{e} d\varphi_{e}$ and
$dO_{2}=\sin\theta_{2} d\theta_{2} d\varphi_{2}$ are the elements of solid angles containing vectors $\mathbf{p}$ and
$\mathbf{k}_{2}$ ($\mathbf{p}$ is the asymptotic value of electron momentum at infinity). The amplitudes $\mathfrak{M}_{atto}^{1,2}(\mathbf{p})$ are given by\footnote{Derivation of the laser-assisted attosecond
amplitudes follows the procedure
\cite{Yudin2007,Yudin2008}. We use the Fourier representation (\ref{atto-pulse-2}) and Jacobi-Anger expansion \cite{Jacobi1836,Anger1855} (see also, e.g., \cite{Watson1922,Gradsteyn2000}).}.
\begin{eqnarray} \label{la-ampl-comp}
\mathfrak{M}_{atto}^{(1,2)}(\mathbf{p})&=& -i D \sum_{m,
n=-\infty}^{\infty}R(\omega^{(1,2)}) \mathfrak{M}_{fi}^{(1,2)}(\mathbf{p}) e^{-in\varphi_{0}}
\nonumber \\ &\times&  (-1)^{m} i^{n}
J_{m}(\mathrsfs{M}_{1}-\mathrsfs{M}_{2}) J_{n}(\mathrsfs{N}),
\end{eqnarray}
where the Compton ionization amplitude $\mathfrak{M}_{fi}^{(2)}(\mathbf{p})$ was given above, see formula (\ref{ampl-comp}), and the monochromatic photoionization amplitude is
\begin{eqnarray} \label{ampl-photo}
\mathfrak{M}_{fi}^{(1)} (\mathbf{p})= \frac{i}{\sqrt{\Omega_{1}'}} \langle f \mid
e^{i \mathbf{k}_{1}'\mathbf{r}} \mathbf{e}_{1} \cdot \mathbf{\nabla} \mid i\rangle.
\end{eqnarray}
The per-channel frequency defects for photo- and Compton ionization are respectively
\begin{eqnarray} \label{omega-laphoto}
\omega^{(1)} =\frac{p^{2}}{2}+I_{p}-\Omega_{1}
+[2(\mathrsfs{M}_{1}+\mathrsfs{M}_{2}+m)+n]\omega _{0},
\end{eqnarray}
\begin{eqnarray} \label{omega-las}
\omega^{(2)} = \omega_{1}
+[2(\mathrsfs{M}_{1}+\mathrsfs{M}_{2}+m)+n]\omega _{0}.
\end{eqnarray}
The argument of Bessel functions $J_m(x)$ in equation
(\ref{la-ampl-comp}) depends on  $\mathrsfs{M}_{1,2}=E_{1,2}^{2}/(2
\omega _{0})^{3}$. In the geometry, where two parameters
$\mathrsfs{N}_{1,2}=\mathbf{p} \cdot
\mathbf{E}_{1,2}/\omega_{0}^{2}$ are not zero simultaneously, the angle $\varphi_{0}$ is defined by the equations
$\cos{\varphi_{0}}=\mathrsfs{N}_{1}/\mathrsfs{N}$ and
$\sin{\varphi_{0}}=-\mathrsfs{N}_{2}/\mathrsfs{N}$, where
$\mathrsfs{N}=\sqrt{\mathrsfs{N}_{1}^{2}+\mathrsfs{N}_{2}^{2}}$;
at $\mathrsfs{N}_{1,2}\rightarrow 0$ the sum on $n$ in
(\ref{la-ampl-comp}) tends to unity.
\section{Applications}
There are many possibilities for measurement because Compton spectra depend on the laser polarization and Compton scattering geometry. We illustrate the general result (\ref{spectrum-6}) in the case of a linearly polarized laser field: $\mathbf{E}_{L}= \mathbf{E}_{1}$ and $\mathbf{E}_{2}=0$ in equation (\ref{vect-pot-las}); the laser wavelength is 800 nm. We consider Compton ionization by an isolated attosecond
x-ray pulse and by two replicas of the same pulse. A coherent attosecond source of photons with the energy of $\approx$8 keV is becoming available at the Linc Coherent Light Source \cite{SLAC2002}. For an illustration of the general expressions, we confine ourselves to this experimental setup and consider the non-relativistic regime where the central frequency of incident photon is $\Omega_{1}$=8 keV. The hydrogenic $1s$ state is chosen as the initial electronic bound state. In the following, we present Compton electron spectra as well as Compton lines in units of $S_{0}=[(\mathbf{e}_{1}\cdot \mathbf{e}_{2})A_{0}/V]^{2}$.
\subsection{Compton electron spectra}

The difficulty in measuring the temporal structure of x-ray pulses is mainly due to the lack of efficient nonlinear processes that are scalable to shorter wavelength. The x-ray photo- and Compton ionization of atoms in the presence of a strong laser field circumvents this problem. The laser field then controls the motion of the electron in the continuum. In the x-ray ionization process, a strong laser electric field acts as an ultrafast phase modulator for the electron wavepackets. This effect can be used to characterize the electron wavepackets and the ionizing x-ray fields. When the x-ray pulse duration falls below one period of the laser field, one can induce a time-dependent acceleration of electrons (i.e. streaking) or a time-independent acceleration (i.e. spectral shearing of photoelectrons), depending on the phase of the laser field with respect to the attosecond x-ray pulses. Both effects give opportunities to measure attosecond x-ray pulses. With the photoelectron spectral shearing technique, one can perform spectral phase interferometry of two electronic wavepackets ionized by two identical but time-delayed x-ray pulses. This method makes it possible to retrieve the spectral phase of the electron wavepackets and, thus, of the ionizing x-ray pulse.

In the hard x-ray scattering regime, due to approximate momentum conservation
$\mathbf{p}\approx \mathbf{k}_{1}'-\mathbf{k}_{2}$. Therefore, the outgoing continuum electron will preferably appear along $\mathbf{k}_{1}'-\mathbf{k}_{2}$. Accordingly, we choose the incoming x-ray momentum $\mathbf{k}_{1}'$ along the Cartesian $z$ axis and use the scattering geometry where
$\theta_{e}=10^{\circ}$, $\theta_{2}=180^{\circ}$ and $\varphi_{e}=\varphi_{2}=0^{\circ}$.

In figures 2 and 3 we report the energy-resolved spectra of
Compton electrons at the laser field intensity
$I=2\times 10^{12}$ Wcm$^{-2}$. The scattered photon energy
is fixed, $\Omega_{2}$=7757 eV. This energy has been chosen in accordance with the value of hydrogen ionization potential and the Compton's formulae \cite{Compton1923}
\begin{eqnarray} \label{Compton-free}
\Omega_{2}=\Omega_{1}/(1+\mu) \approx (1- \mu )\Omega_{1},
\quad
E_{e}\approx \mu(1-\mu) \Omega_{1},
\end{eqnarray}
where $\mu=(\Omega_{1}/c^{2})(1-\cos{\theta})$ and $\theta$ is the angle between $\mathbf{k}_{1}'$ and $\mathbf{k}_{2}$.

As in the photoelectron ASC, the strong laser field distorts the Compton electron spectrum on a sub-cycle time scale. Figure 2 shows the Compton electron spectra which are produced by a single x-ray pulses of 100 asec FWHM and different chirps.
Times of the pulse maxima are $t_{1}$=0
(curves 1a, 1b and 1c) and $t_{1}=\pi/2\omega_{0}$ (curves 2a, 2b and 2bc). The chirps are: $\xi$=0 (curves 1a and 2a), $\xi$=3 (curves 1b and 2b) and $\xi$=-3 (curves 1c and 2bc). One can observe the dependence of the position of spectral maximum and its amplitude on $t_{1}$ and the chirp. The maximum streaking speed occurs at the time where $\mathbf{E}_{1}\sin (\omega_{0}t)$ is at an extremum. It is reflected in dependence of the spectra on the sign of the chirp: in the case of $t_{1}$=0, at different signs of the chirp, we can clearly see the difference between spectra (curves 1b and 1c), while the spectra calculated at
$t_{1}=\pi/2\omega_{0}$ and $\xi=\pm$3 are identical (curve 2bc).

The energy shift between electron wavepackets, needed for complete phase measurements, can be induced by laser-assisted
Compton ionization. To demonstrate the Compton ASPI, we show (figure 3) the Compton electron spectra generated by two replicas of an attosecond pulse, each with 100 asec FWHM. Times of the pulse maxima are $t_{1}$=0 and $t_{2}=\pi/2\omega_{0}$. Owing to time separation of $\pi/2\omega_{0}$ between the replicas, interference is observed in the spectral fringes. Its periodicity
depends slightly on the chirp and Compton electron energy,
as is illustrated in figure 3(b). When the pulse is
positively chirped, the fringes shift to lower energies on the low-energy side of the spectrum, and to higher energies on the high-energy side. To reconstruct the spectral phase from the Compton electron spectrum, one would use the same algorithm as in the photoelectron ASPI \cite{Quere2003,Itatani2004}.

For comparison, in figure 4 we show atomic energy-resolved Compton electron spectra generated by longer x-ray pulses at different laser field intensities. Spectra shown in figure 4(a) are generated by a single chirped ($\xi$=3) attosecond ($\tau$=300 asec) pulse arriving at $t_{1}$=0. Spectra shown in figure 4(b) are generated by two replicas of chirped ($\xi$=-3) attosecond ($\tau$=300 asec) pulse arriving at $t_{1}$=0 and $t_{2}=\pi/2\omega_{0}$. In both cases, curves 1--4 correspond to laser-free Compton ionization (1) and  laser intensities $I_{2}=3\times 10^{11}$ Wcm$^{-2}$,
$I_{3}=9\times10^{11}$ Wcm$^{-2}$ and $I_{4}=4\times10^{12}$
Wcm$^{-2}$ respectively. As in atomic attosecond photoionization \cite{Yudin2007,Yudin2008}, it is possible to distinguish five mechanisms contributing to the development of the Compton interference pattern. Due to an interplay between a number of sequential and non-sequential temporal and spatial interference mechanisms, the shapes and positions of Compton electron peaks are strongly dependent on the relative orientation of field polarization directions and other parameters of the fields. Positions of the individual minima and maxima can be measured accurately and can be used to characterize both fields, at least in principle.
\subsection{Attosecond Compton lines}
We now consider the energy-resolved spectra of scattered photons, the Compton lines:
\begin{eqnarray} \label{spectrum-3}
\frac{d^{3}P_{Comp}}{d\Omega_{2}dO_{2}}\equiv S_{Comp}^{(3)} =\int
S_{Comp}^{(6)} dpdO_{e}.
\end{eqnarray}
Note that matrix elements, containing the operators $\hat{G}_{1,2}$ in the amplitude (\ref{ampl-comp}), can be omitted in the Compton line, since their contribution to the integral (\ref{spectrum-3}) is a second-order correction. The first-order relativistic corrections to the Compton line arise from the operator $\hat{G}_{0}$.

To take integral (\ref{spectrum-3}) numerically, we used 29th-order Lebedev grid \cite{Lebedev1977} for the angular part and equidistant momentum grid with the spacing of $\approx1.5\times10^{-3}$ a.u. We verified convergence of the results by repeating selected integrals with the 47th-order Lebedev grid \cite{Lebedev1992} and $\approx 7.5\times10^{-4}$ a.u. momentum spacing.

Integration in formula (\ref{spectrum-3}) leads to the following result: for experimentally realizable x-ray durations \textit{symmetric normalized} Compton lines almost coincide with the laser-free Compton lines for the same frequencies $\Omega_{1,2}$ and scattering angles $\theta$ (in our case, at $I \lesssim 10^{13}$ Wcm$^{-2}$). Furthermore, the initial x-ray parameters such as chirp and duration play a minor role at these moderate laser-field intensities. The question immediately suggested itself: is there a regime where the differential probability (\ref{spectrum-3}) is directly proportional to the monochromatic Compton profile $d^{3}\sigma_{CP}/d\Omega_{2}dO_{2}$ \cite{Eisenberger1970}? This situation is realized when (i) the oscillations inside x-ray pulse are  quasi-monochromatic and the laser fields are of moderate intensity; and simultaneously (ii) the width of the Compton line due to the initial electronic \textit{bound state} momentum distribution [$\Delta \Omega_{2}^{(b)} \sim
(Z_{a}/c)\Omega_{1}\sin(\theta/2)$] is much broader then the bandwidth of the initial attosecond photon field [$\Delta \Omega_{2}^{(atto)} \sim \Delta\Omega_{1}^{(atto)}\sim 1/\tau$],
\begin{eqnarray} \label{formula1}
\tau Z_{a}\Omega_{1}\sin(\theta/2)\gg c;
\end{eqnarray}
and (iii) the main contribution to the scattering comes from the ionization channel. Using the results obtained for monochromatic Compton line spectra \cite{Kaplan1975}, one can find that, for the 1s initial electron state, with the accuracy which is higher than experimental resolution, the last criterion becomes
\begin{eqnarray} \label{formula2}
\hbar \Omega_{1} \sin{(\theta /2)} \gtrapprox 6 Z_{a} keV.
\end{eqnarray}
Then the \textit{attosecond} Compton lines (\ref{spectrum-3}) are given by
\begin{eqnarray} \label{CProfiles}
S_{Comp}^{(3)} \propto \frac{d^{3}\sigma_{CP}}{d\Omega_{2}dO_{2}}=
\frac{\Omega_{2}}{\Omega_{1}} \frac{(\mathbf{e}_{1}\cdot \mathbf{e}_{2})^{2}}{(2\pi)^{3}c^{4}} \nonumber  \\
\times \int |\Phi_{i}(\mathbf{p})|^{2}  \delta
(\Omega_{1}-\Omega_{2}-\frac{k^{2}}{2}-\mathbf{k}\cdot\mathbf{p})d^{3}p,
\end{eqnarray}
whereas $|\Phi_{i}(\mathbf{p})|^{2}$ is the initial electron momentum density. It should be emphasized that, as in the monochromatic case (details can be found in \cite{Dykhne1996}), the accuracy of the remarkable result for the Compton profile $d^{3}\sigma_{CP}/d\Omega_{2}dO_{2}$ in formula (\ref{CProfiles}) is
\begin{eqnarray} \label{formula3}
\delta \sim \frac{I_{p}}{\Omega_{1}} \Bigl(\frac{I_{p}}{\Omega_{1}}+\frac{Z_{a}}{c}\Bigr).
\end{eqnarray}

The situation changes drastically in the presence of an intense laser field.  In figure 5 we present \textit{asymmetric Compton lines} $S^{(3)}_{Comp}$ calculated at a relatively high laser-field intensity $I=1\times 10^{14}$ Wcm$^{-2}$. Spectra in  figure 5(a) are generated by a transform-limited ($\xi =0$) attosecond x-ray pulse with the duration $\tau=1200$ asec ($\tau_{FWHM}\approx 2000$ asec). In figure 5(b) we use a chirped ($\xi=-3$) attosecond x-ray pulse with the duration $\tau=300$ asec ($\tau_{FWHM}\approx 1600$ asec).  Spectra 1 and 2 in figure 5 are produced by a single chirped attosecond  pulse arriving at $t_{1}$=0. Spectra 3 and 4 are generated by two replicas at $t_{1}$=0 and
$t_{2}=\pi/2\omega_{0}$ accordingly. Curves 1 and 3 correspond to photon scattering angle $\theta$=120$^{\circ}$, while curves 2 and 4 correspond to $\theta$=180$^{\circ}$.

Let us examine the origin of the asymmetry of attosecond Compton lines. From figure 4 we observe that the maximum of the Compton electron energy in the absence of the laser field is $E_e \approx 230$ eV. This value of $E_e$ corresponds to $m=n=0$ in equation (\ref{la-ampl-comp}). Therefore, this is the boundary between processes of stimulated absorption ($m,n > 0$) and emission ($m,n < 0$). The boundary appears at $\hbar \Omega_2 \approx 7.77$ keV in the spectra of scatted photons. Mathematically speaking, the form of the ``Volkov part'' of the separable Coulomb-Volkov wavefunction
\begin{eqnarray} \label{Volkov}
\Psi_{V}(\mathbf{r}, t) \propto
\exp{\Bigl\{-\frac{i}{2}\int^{t}[\mathbf{p}+\mathbf{A}_{L}(t)]^{2}
dt \Bigr\}}
\end{eqnarray}
is the root of the asymmetry. After expanding $\Psi_{V}(\mathbf{r}, t)$ into a series of Bessel functions, we note that the effective number of channels of absorption as well as their contribution to the amplitude of laser-assisted Compton ionization is larger than the number and contribution from emission channels. At sufficiently high laser-field intensities the difference between stimulated absorption and emission becomes clearly noticeable: $\gtrsim 80$ photons ($\gtrsim 120$ eV) of the laser field are absorbed, and $\lesssim 60$ photons ($\lesssim 90$ eV) of the laser field are emitted. It is remarkable that the asymmetry depends on positions of the peaks. Note that in spite of the slight difference between the durations of the pulses in figures 5(a) and 5(b), the spectra differ tremendously for the case of two replicas. However, the spectra generated by a single pulse with the peak at $t_1 =0$ deviate quantitatively while retaining similar shapes. Therefore, the quantitative and qualitative difference in the case of the two replicas is determined by the replica with the maximum at $t_2 = \pi/2\omega_0$.

Unexpectedly, at high laser-field intensities attosecond interference exhibits substantial dependence on the relative x-ray laser-field phase and x-ray pulse chirp not only in differential Compton electron spectra (figures 2--4), but even in Compton lines, which are integrated over the electron momenta (figure 5).
\section{Conclusion}
We extend and improve the previous procedure \cite{Itatani2002,Kitzler2002,Quere2003,Itatani2004,Yudin2007,Yudin2008}
in a \textit{completely ab initio manner} and without recourse a semiclassical treatment, by using the potential of the initial attosecond photon state. We thus develop a rigorous semi-relativistic analytical theory of laser-assisted attosecond Compton ionization. Strong-laser electric field acts as an ultrafast phase modulator for the Compton electron wavepackets. The theory is applicable to arbitrary laser-field polarizations and x-ray pulse parameters. If the x-ray pulse duration falls below one period of the laser field, two new methods for measuring hard x-ray pulse durations---attosecond Compton streak camera and spectral phase interferometry---become feasible. Both techniques require coincidence measurement of the scattered photon and Compton electron. If the scattered x-ray pulse is accurately measured, then similar to the photoelectron ASPI, there are no fundamental limits to the resolution of attosecond Compton scattering measurements.

Once the x-ray pulse is characterized using a well-understood atom such as hydrogen, a new frontier of attosecond science will be to utilize the Compton streak camera to measure unknown dynamics in target systems---nuclei, atoms, molecules or solids---just as photoelectron attosecond streak camera is used to measure dynamics initiated by attosecond XUV pulses or by recollision \cite{Corkum2007}.

The relative cross sections of photoionization and Compton scattering are sensitive to the incident photon energy and ionization potential. In the case of a weakly bound electron, the Compton channel can be identified due to relatively low energies of Compton electrons: $E_{e}^{(Comp)}\ll E_{e}^{(photo)}$. In the delicate case of laser-assisted ionization, Compton ``slow''-electron spectra can be resolved with much higher accuracy than for the ``high''-energy photoelectron. Therefore, in semi-relativistic hard x-ray low-$Z_{a}$ regime, Compton probe spectroscopy remains feasible, while photoionization probe spectroscopy would require experimentally unattainable resolution to yield useful information on the attosecond dynamics of the target.

In the hard x-ray regime, even an ultrashort pulse comprises many oscillations of electromagnetic field and shows many similarities with monochromatic radiation. As a result, attosecond Compton scattering permits accurate imaging of electronic structure, without sacrificing either electron momentum or the time resolution. Once attosecond
hard x-ray pulses synchronized to infrared laser fields become experimentally available, we envision laser-assisted Compton scattering becoming an indispensable technique in studies of fast processes.

Finally, we demonstrate that treatment of the x-ray field quantization is essential for characterization of attosecond x-ray pulses through photoionization processes as well. The photoionization probability (\ref{spectrum-photo}) depends on $\omega^{(1)}$ through the renormalized incident photon energy $\Omega_{1}'$, see formula (\ref{ampl-photo}). This dependence, which was not previously recognized in the published literature, becomes important at very short durations of the x-ray pulse. It gives relative contribution of the order $\omega^{(1)} / \Omega_{1} \sim (\Omega_{1} \tau)^{-1}$ to the photoionization probability, and can critically affect the delicate dependence of the photoelectron spectra on the the x-ray pulse and laser-field parameters. Chirp-dependent attosecond interferences in the Coulomb-Volkov continuum \cite{Yudin2007,Yudin2008} are especially sensitive to the new effect. As an example, for the parameters considered in \cite{Sansone2009} ($\tau_{FWHM}\approx$50 asec and $\Omega_{1}\approx$100 eV) $(\Omega_{1} \tau)^{-1}$ is approximately 20$\%$, and cannot be neglected.
\section*{Acknowledgments}
We appreciate the valuable discussion with A~A~Zholents. This work has been partially supported by NSERC Grant SRO  5796-299409/03.

{\bf Supplementary material:} Fortran implementation of differential Compton electron spectra and Compton lines.

\begin{figure}
	\begin{center}
		\includegraphics[scale=0.5]{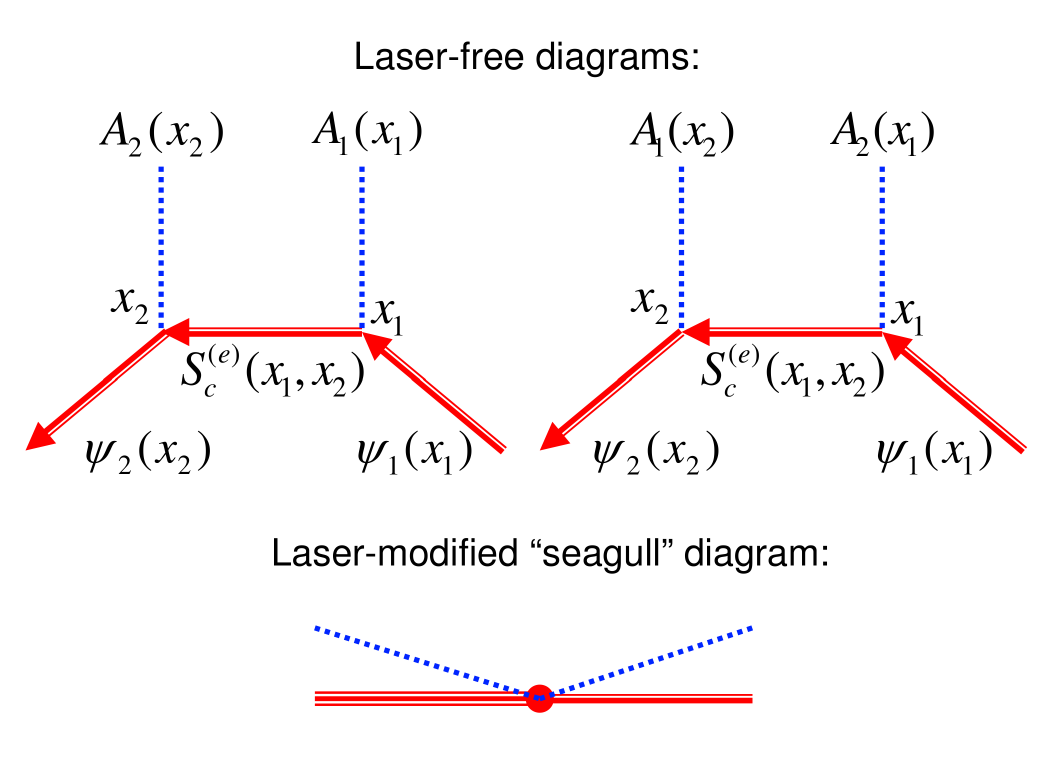}
		\caption{Compton ionization diagrams included in our treatment. List of notations, used in
    		\textit{laser-free} diagrams, are in the text.
    		Triple electron line in laser-modified ``seagull'' diagram describes the electron state in the separable Coulomb-Volkov continuum.}
	\end{center}
\end{figure}

\begin{figure}
	\begin{center}
		\includegraphics[scale=0.5]{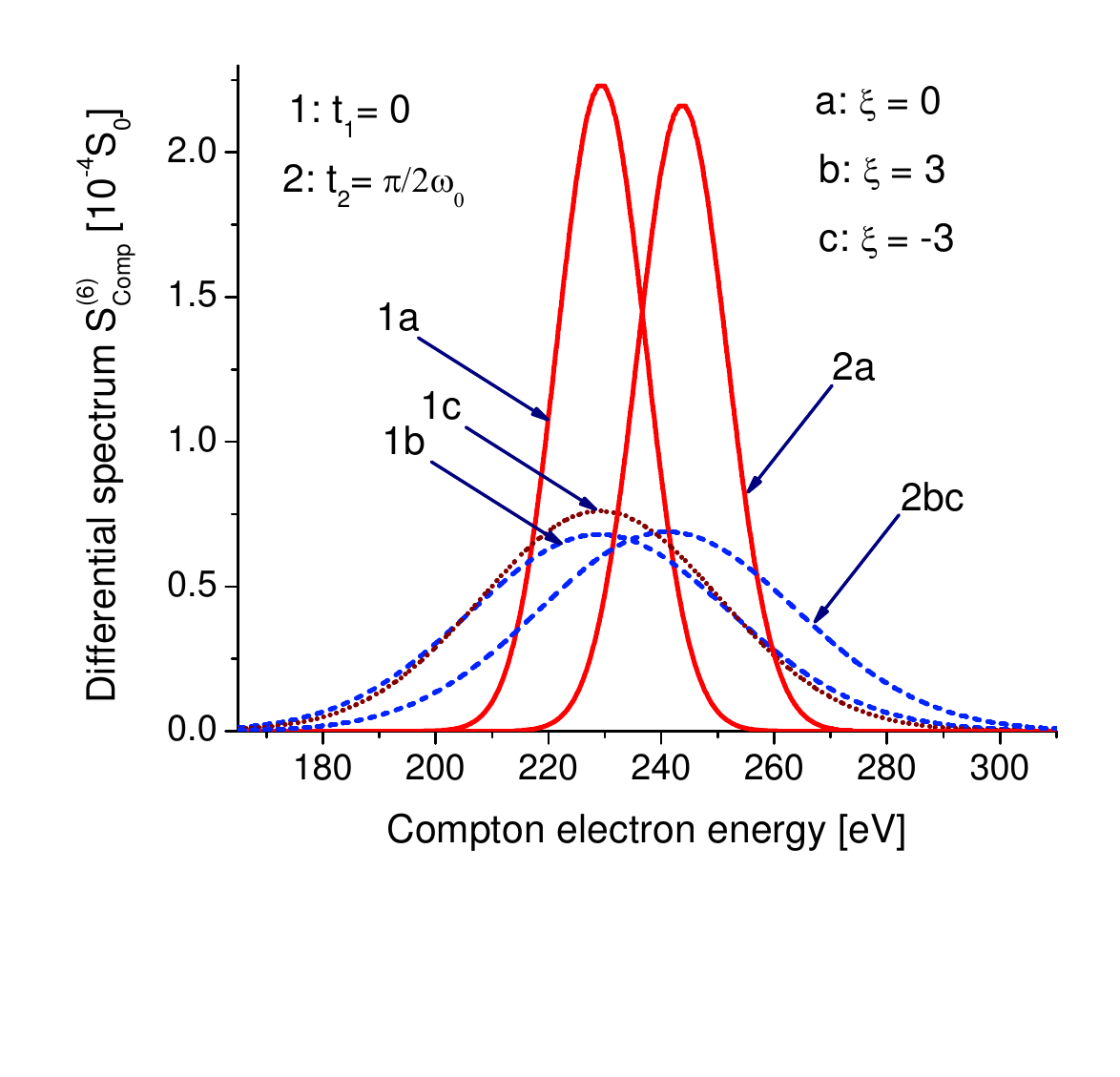}
		\caption{\textit{Compton streak camera.} Electron spectra are produced by a single x-ray pulses of 100 asec FWHM and different chirps. The central incident photon energy and the scattered photon energy are $\Omega_{1}$=8 keV and $\Omega_{2}$=7757 eV. The laser wavelength and intensity are 800 nm and $I=2\times 10^{12}$ Wcm$^{-2}$. Times of the pulse maxima are $t_{1}$=0 (curves 1) and $t_{1}=\pi/2\omega_{0}$ (curves 2). The value $S_{0}$, scattering geometry and plot labels are defined in the text.}
	\end{center}
\end{figure}

\begin{figure}
	\begin{center}
		\includegraphics[scale=1]{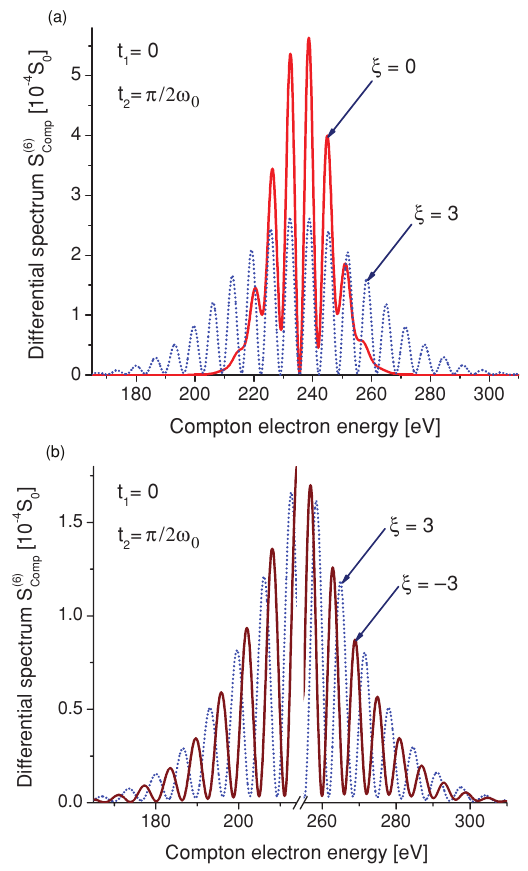}
		\caption{\textit{Compton spectral phase interferometry.} The incident and scattered photon energies, as well as scattering geometry, laser-field parameters and $S_{0}$ are the same as in figure 2. Electron spectra are generated by two replicas of an attosecond pulse of 100 asec FWHM. Times of two pulse maxima are $t_{1}$=0 and $t_{2}=\pi/2\omega_{0}$. Chirp dependence is discussed in the text.}
	\end{center}
\end{figure}

\begin{figure}
	\begin{center}
		\includegraphics[scale=1]{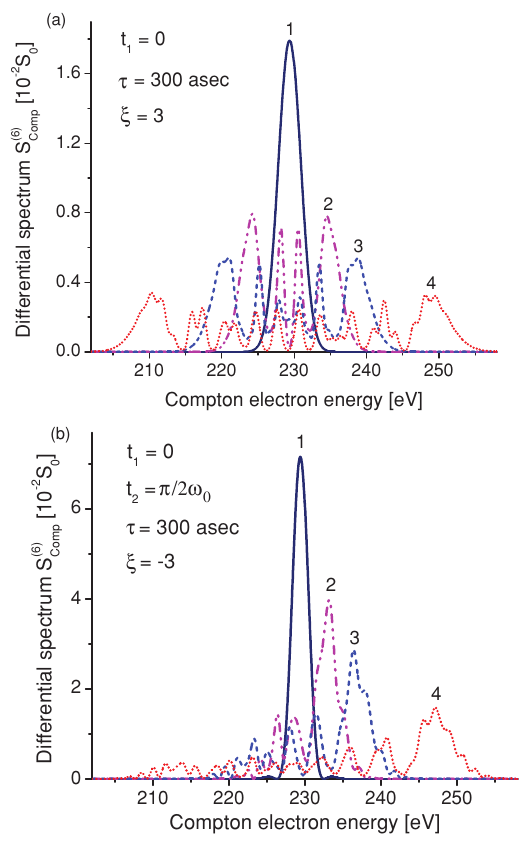}
		\caption{Energy-resolved Compton electron spectra generated by a single chirped attosecond pulse and by two replicas of a single chirped attosecond pulse. The incident and scattered photon energies, as well as scattering geometry, and $S_{0}$ are the same as in figure 2. Curves 1--4 correspond to the laser-free case (1) and the laser intensities $I_{2}=3\times 10^{11}$ Wcm$^{-2}$, $I_{3}=9\times 10^{11}$ Wcm$^{-2}$ and $I_{4}=4\times 10^{12}$ Wcm$^{-2}$ respectively. Other parameters are defined in the text.}
	\end{center}
\end{figure}

\begin{figure}
	\begin{center}
		\includegraphics[scale=1]{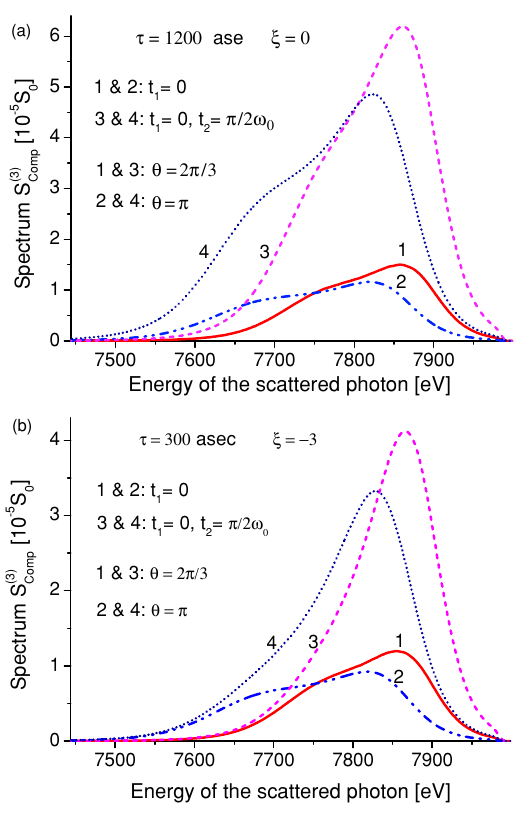}
		\caption{Compton lines [in units of $S_{0}$] at the laser field intensity $I=1\times10^{14}$ W/cm$^{2}$. Spectra 1 and 2 are produced by a single (transform-limited or chirped) attosecond  pulse arriving at $t_{1}$=0, while spectra 3 and 4 are generated by two replicas at $t_{1}$=0 and $t_{2}=\pi/2\omega_{0}$. Curves 1 and 3 correspond to photon scattering angle $\theta$=120$^{\circ}$, while curves 2 and 4  correspond to $\theta$=180$^{\circ}$. Chirps and x-ray pulse durations in figures 5(a) and 5(b) are different and are defined in the text.}
	\end{center}
\end{figure}

%

%

%


\section*{References}
\begin{thebibliography}{99}
%

\bibitem{Paul2001} Paul~P~M, Toma~E~S, Berger~P, Mullot~G, Auge~F, Balcou Ph, Muller~H~G and Agostini~P 2001 \textit{Science} \textbf{292} 1689

\bibitem{Hentschel2001} Hentschel~M, Kienberger~R, Spielmann~Ch, Reider~G~A, Milosevic~N, Brabec~T, Corkum~P, Heinzmann~U, Drescher~M and Krausz~F 2001 \textit{Nature} \textbf{414} 509

\bibitem{Itatani2002} Itatani~J, Qu\'er\'e~F, Yudin~G~L, Ivanov~M~Yu, Krausz~F and Corkum~P 2002 \textit{Phys. Rev. Lett.} \textbf{88} 173903

\bibitem{Kitzler2002} Kitzler~M, Milosevic~N, Scrinzi~A, Krausz~F and Brabec~T 2002 \textit{Phys. Rev. Lett.} \textbf{88} 173904

\bibitem{Quere2003} Qu\'er\'e F, Itatani~J, Yudin~G~L and Corkum~P~B 2003 \textit{Phys. Rev. Lett.} \textbf{90} 073902

\bibitem{Itatani2004} Itatani~J, Qu\'er\'e F, Yudin~G~L and Corkum~P~B 2004 \textit{Laser Phys.} \textbf{14} 344

\bibitem{Corkum2007} Corkum~P~B and Krausz~F 2007 \textit{Nat. Phys.} \textbf{3} 381

\bibitem{Krausz2009} Krausz~F and Ivanov~M 2009 \textit{Rev. Mod. Phys.} \textbf{81} 163

\bibitem{Nisoli2009} Nisoli~M and Sansone~G 2009 \textit{Prog. Quantum Electron.} \textbf{33} 17

\bibitem{Saldin2002} Saldin~E~L, Schneidmiller~E~A and Yurkov~M~V 2002 \textit{Opt. Commun.} \textbf{212} 377

  Saldin~E~L, Schneidmiller~E~A and Yurkov~M~V 2004 \textit{Opt. Commun.} \textbf{237} 153

  Saldin~E~L, Schneidmiller~E~A and Yurkov~M~V 2004  \textit{Opt. Commun.} \textbf{239} 161

  Saldin~E~L, Schneidmiller~E~A and Yurkov~M~V 2006 \textit{Phys. Rev. ST Accel. Beams} \textbf{9} 050702

\bibitem{Zholents2004} Zholents~A~A and Fawley~W~M 2004 \textit{Phys.~Rev.~Lett.} \textbf{92} 224801

\bibitem{Zholents2005a} Zholents~A~A 2005 \textit{Phys. Rev. ST Accel. Beams} \textbf{8} 040701

\bibitem{Zholents2005b} Zholents~A~A and Penn~G 2005 \textit{Phys. Rev. ST Accel. Beams} \textbf{8} 050704

\bibitem{Zholents2008} Zholents~A~A and Zolotorev~M~S 2008 \textit{New J. Phys.} \textbf{10} 025005

\bibitem{Ding2009} Ding~Y, Huang~Z, Ratner~D, Bucksbaum~P and Merdji~H 2009 \textit{Phys. Rev. ST Accel. Beams} \textbf{12} 060703

\bibitem{Kim2009} Kim~D, Lee~H, Chung~S and Lee~L 2009 \textit{New J. Phys.} \textbf{11} 063050

\bibitem{Bethe1957} Bethe~H~A and Salpeter~E~E 2008 \textit{Quantum Mecanics of One- and Two-Electron Atoms} (New York: Dover)

\bibitem{Compton1923} Compton~A~H 1923 \textit{Phys.~Rev.} \textbf{21} 207, 483

\bibitem{Du Mond1929} Du Mond~J~W~M 1929 \textit{Phys.~Rev.} \textbf{33} 643; 1930 \textit{Rev.~Mod.~Phys.} \textbf{5} 1

\bibitem{Eisenberger1970} Eisenberger~P and Platzman~P~M 1970 \textit{Phys. Rev.} A \textbf{2} 415

\bibitem{Cooper2004} Cooper~M~J, Mijnarends~P~E, Shiorani~N, Sakai~N and Bansil~A (eds) 2004 \textit{X-Ray Compton Scattering} (Oxford: Oxford University Press)

\bibitem{Schulke2007} Sch\"{u}lke~W 2007 \textit{Electron Dynamics by Inelastic X-ray Scattering} (Oxford: Oxford University Press)

\bibitem{Akhiezer1969} Akhiezer~A~I and Berestetskii~V~B 1969 \textit{Quantum Electrodynamics} 3rd edn (Moscow: Nauka)

\bibitem{Yudin2007}Yudin~G~L, Patchkovskii~S, Corkum~P~B and Bandrauk~A~D 2007 \textit{J. Phys. B: At. Mol. Opt. Phys.} \textbf{40} F93, 2223

\bibitem{Yudin2008} Yudin~G~L, Patchkovskii~S and Bandrauk~A~D 2008 \textit{J. Phys. B: At. Mol. Opt. Phys.}  \textbf{41} 045602

\bibitem{Furry1951} Furry~W~H 1951 \textit{Phys. Rev.} \textbf{81} 115

\bibitem{Yudin1976} Yudin~G~L 1976 \textit{Sov. Phys. Dokl.} \textbf{21} 591

\bibitem{Dykhne1996} Dykhne~A~M and Yudin~G~L 1996 \textit{Sudden Perturbations and Quantum Evolution} (Moscow: Usp. Fiz. Nauk)

\bibitem{Jacobi1836} Jacobi C G J 1836 J. Reine Math. \textbf{15} 1

\bibitem{Anger1855} Anger C T 1855 \textit{Neueste Schriften Naturforschenden Ges. Danzig} \textbf{5} 1 (see also http://www.digizeitschriften.de/resolveppn/PPN243919689)

\bibitem{Watson1922} Watson G N 1922 \textit{A Treatise on the Theory of Bessel Functions} (Cambridge: Cambridge University Press)

\bibitem{Gradsteyn2000} Gradsteyn I S and Ryzhik I M 2000 \textit{Tables of Integrals, Series, and Products} 6th edn ed A Jeffrey and D Zwillinger (New York: Academic)

\bibitem{SLAC2002} Arthur~J \textit{et al} 2002 Report SLAC-R-593 (Stanford); see also http://www-ssrl.slac.stanford.edu/lcls/cdr/

\bibitem{Kaplan1975} Kaplan~I~G and Yudin~G~L 1976 \textit{Sov. Phys.-JETP} \textbf{42} 4

\bibitem{Lebedev1977} Lebedev~V~I 1977 {\it Sib. Mat. Zh.} {\bf 18} 132

\bibitem{Lebedev1992} Lebedev~V~I and Skorokhodov~A~L 1992 {\it Russian Sci. Dokl. Math.} {\bf 45} 587

\bibitem{Sansone2009} Sansone~G, Ferrari~F, Vozzi~C, Calegari~F, Stagira~S and Nisoli~M 2009 \textit{J. Phys. B: At. Mol. Opt. Phys.} \textbf{42} 134005

%
\end{thebibliography}
\end{document}